\documentstyle[epsfig]{aipproc}
\def\gsim{\,\lower.25ex\hbox{$\scriptstyle\sim$}\kern-1.30ex%
\raise 0.55ex\hbox{$\scriptstyle >$}\,}
\def\lsim{\,\lower.25ex\hbox{$\scriptstyle\sim$}\kern-1.30ex%
\raise 0.55ex\hbox{$\scriptstyle <$}\,}
\newcommand{\ccbar}{c\overline{c}}
\newcommand{\bbbar}{b\overline{b}}
\newcommand{\dstar}{D^*}

\newcommand{\ra}{\rightarrow}
\newcommand{\picb}{\mbox{pb}^{-1}}

\newcommand{\qsq}{${Q^2}$}
\newcommand{\qsqw}{Q^2}
\newcommand{\gevsq}{$\mbox{GeV}^2$}

\newcommand{\cm}{center of mass}
\newcommand{\nlo}{next to leading order}
\newcommand{\degree}{^{\circ}}

\newcommand{\cms}{center of mass}

\newcommand{\colsing}{color singlet}
\newcommand{\coloct}{color octet}
\newcommand{\colsingmodel}{\colsing\ model}

\newcommand{\cs}{cross section}
\newcommand{\Cs}{Cross section}

\newcommand{\jpsi}{$J/\psi$}

\newcommand{\ftcc}{F^{c}_2}

\begin{document}
\title{ Heavy Flavour Physics at HERA}
    
\author{Beate Naroska\footnote{Contribution to ``Workshop on Heavy Quarks at 
Fixed Target'', Fermi National Accelerator Laboratory, Oct. 10, 1998 }}
\address{University of Hamburg\\ II. Institut f\"ur Experimentalphysik\\
Luruper Chaussee 149\\D 22761 Hamburg\\
E-mail:naroska@mail.desy.de}

\maketitle

\begin{abstract}
New results with increased statistics are presented for heavy flavour 
production at \qsq$\lsim 150\,$\gevsq\ and in the photoproduction limit
\qsq$\ra\,0$.
\Cs s for $\dstar$\ production, $\ftcc$, the gluon density in the proton, 
 and inelastic \jpsi\ production are discussed and 
compared to theoretical calculations. A first measurement of the  
$\bbbar$\ \cs\ is reported.
\end{abstract}

\subsection*{Introduction}

At HERA positrons of 27.5 GeV collide with 820 GeV protons yielding
a \cm\ energy of 300 GeV. Heavy flavours are
predominantly produced in pairs by photon gluon fusion. Charm quark production
is expected to be a factor 200 more abundant than bottom quarks at this energy.
Heavy flavour processes give new opportunities of studying perturbative
QCD at \cm\ energies roughly a factor 10 higher than in fixed target
experiments. 

\begin{figure}[h!] 
\centering
\vspace{-0.5cm}
\epsfig{file=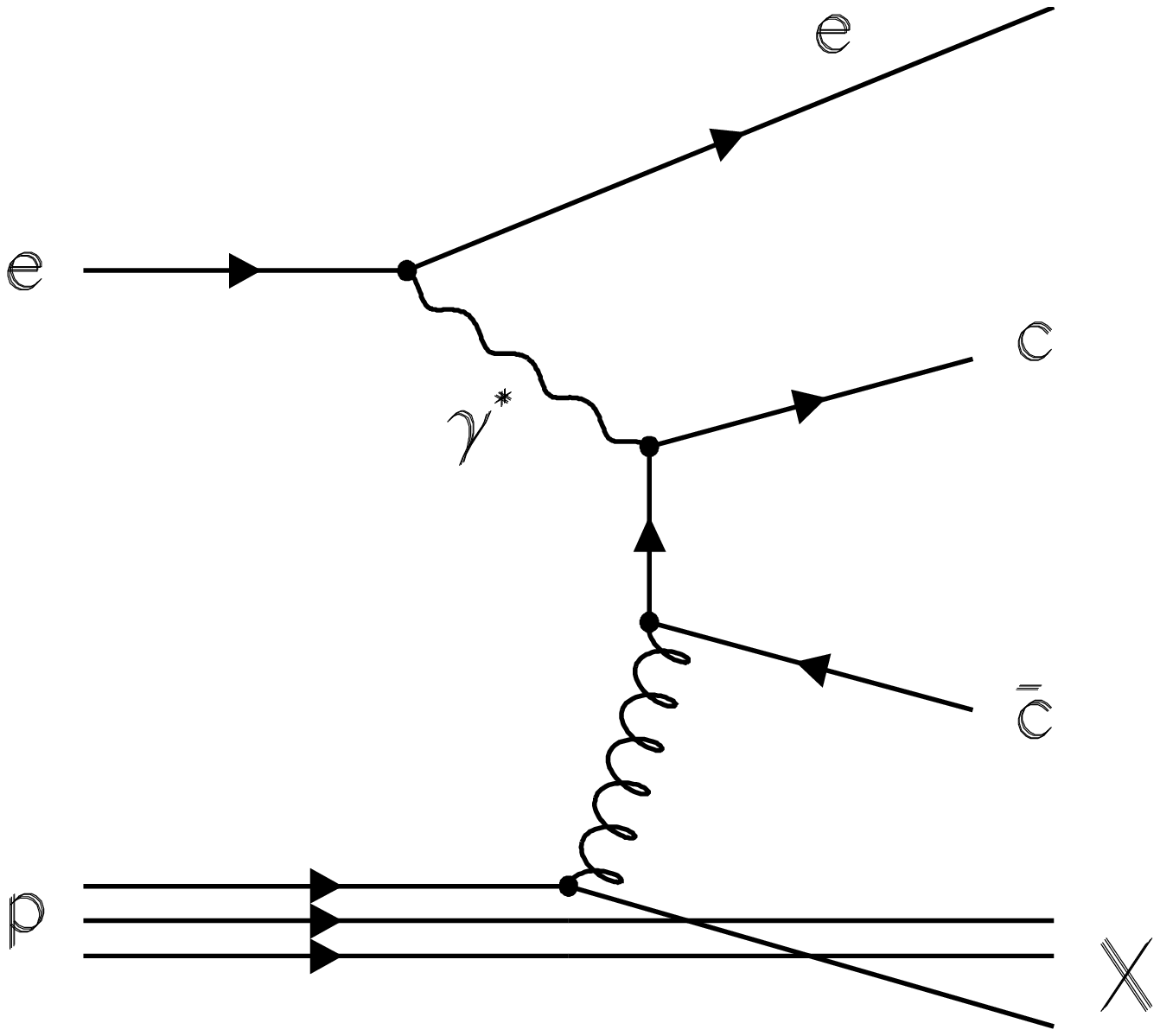,width=6cm,clip=}
\epsfig{file=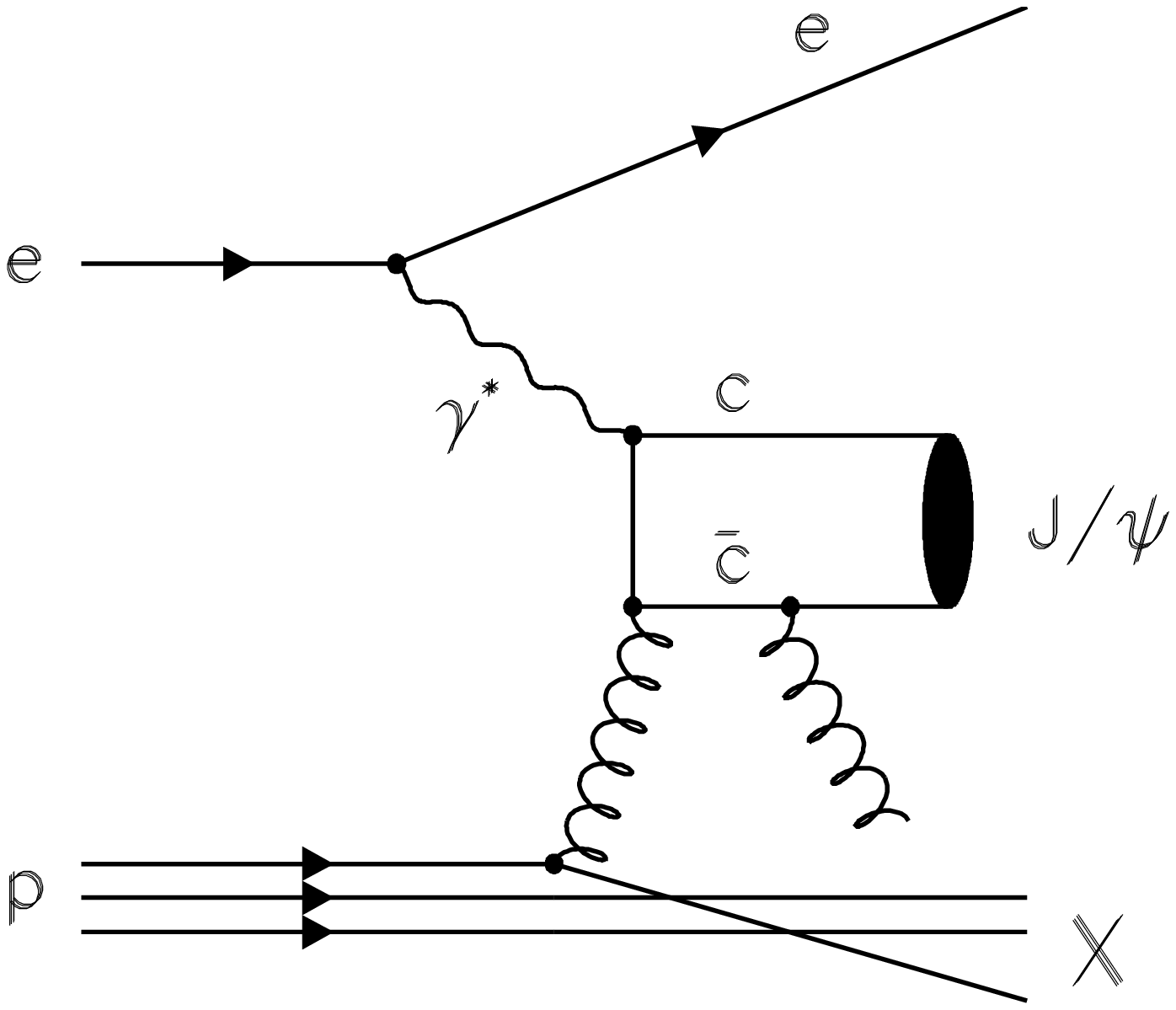,width=6cm,clip=}

\vspace{-1.5cm}
\caption{Charm production via photon gluon fusion (left) and \protect\jpsi\ production in the \colsing\ model.}
\label{fig1}
\end{figure}

There are several methods to tag heavy flavours: 
``Open'' charm production is tagged via reconstruction of $\dstar$ (H1 and ZEUS) or via semi--leptonic decays to electrons (ZEUS). b quarks have been 
measured via semi--muonic decays by H1.
Finally ``hidden'' charm is studied via reconstruction of \jpsi\ (see 
fig.~\ref{fig1}).
$\psi(2s)$\ and $\Upsilon$ Mesons have as yet been reconstructed only in 
diffractive processes~[1a] at HERA and will not be reported here.

\begin{figure}[t!]
\vspace{-0.5cm}
\epsfig{file=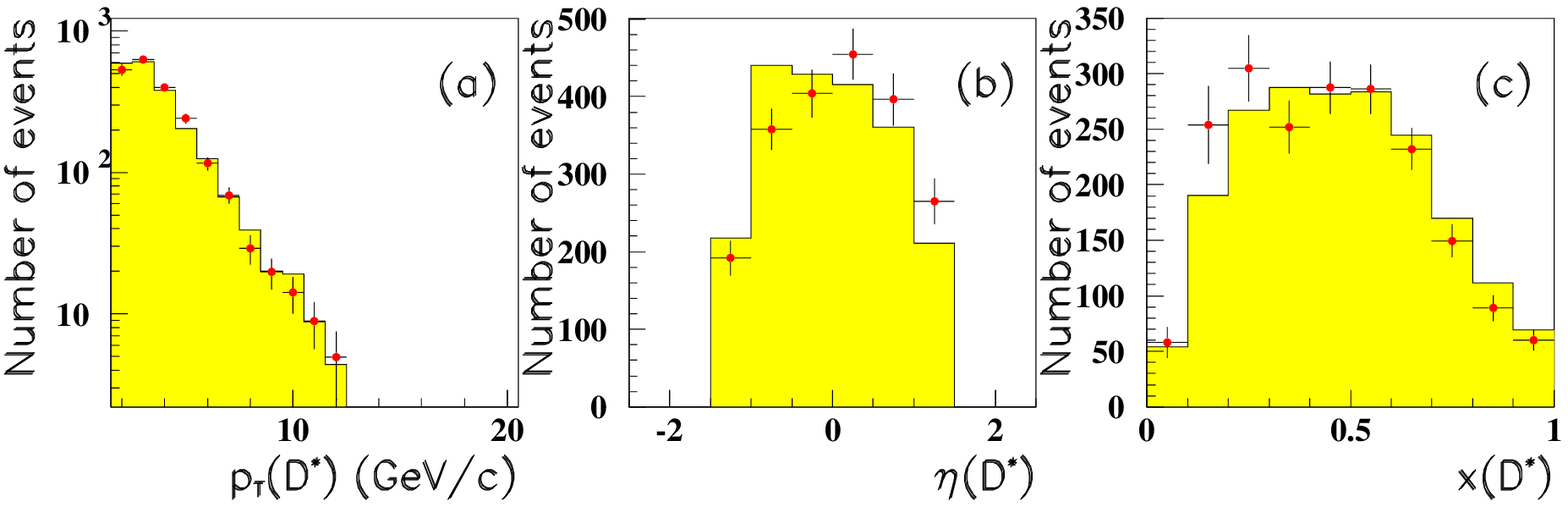,width=15cm,bbllx=0pt,bblly=0pt,bburx=550pt,bbury=190pt,clip=}
\caption {ZEUS: Reconstructed $D^*$ related quantities compared to 
Monte Carlo simulation RAPGAP~\protect\cite{rapgap}.}
\label{fig2}

\vspace{-0.8cm}  
\epsfig{file=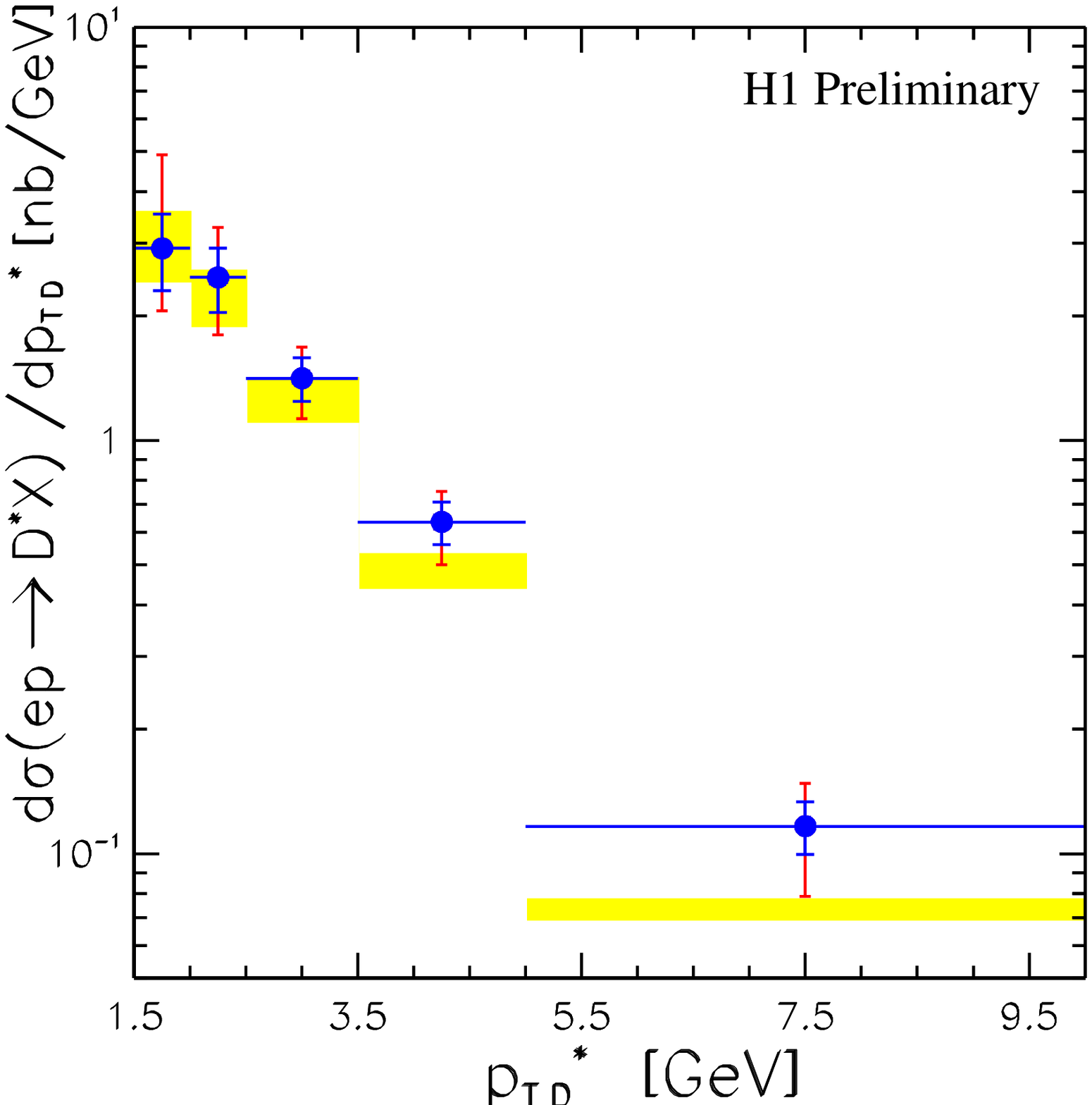,height=5.1cm,clip=}
\epsfig{file=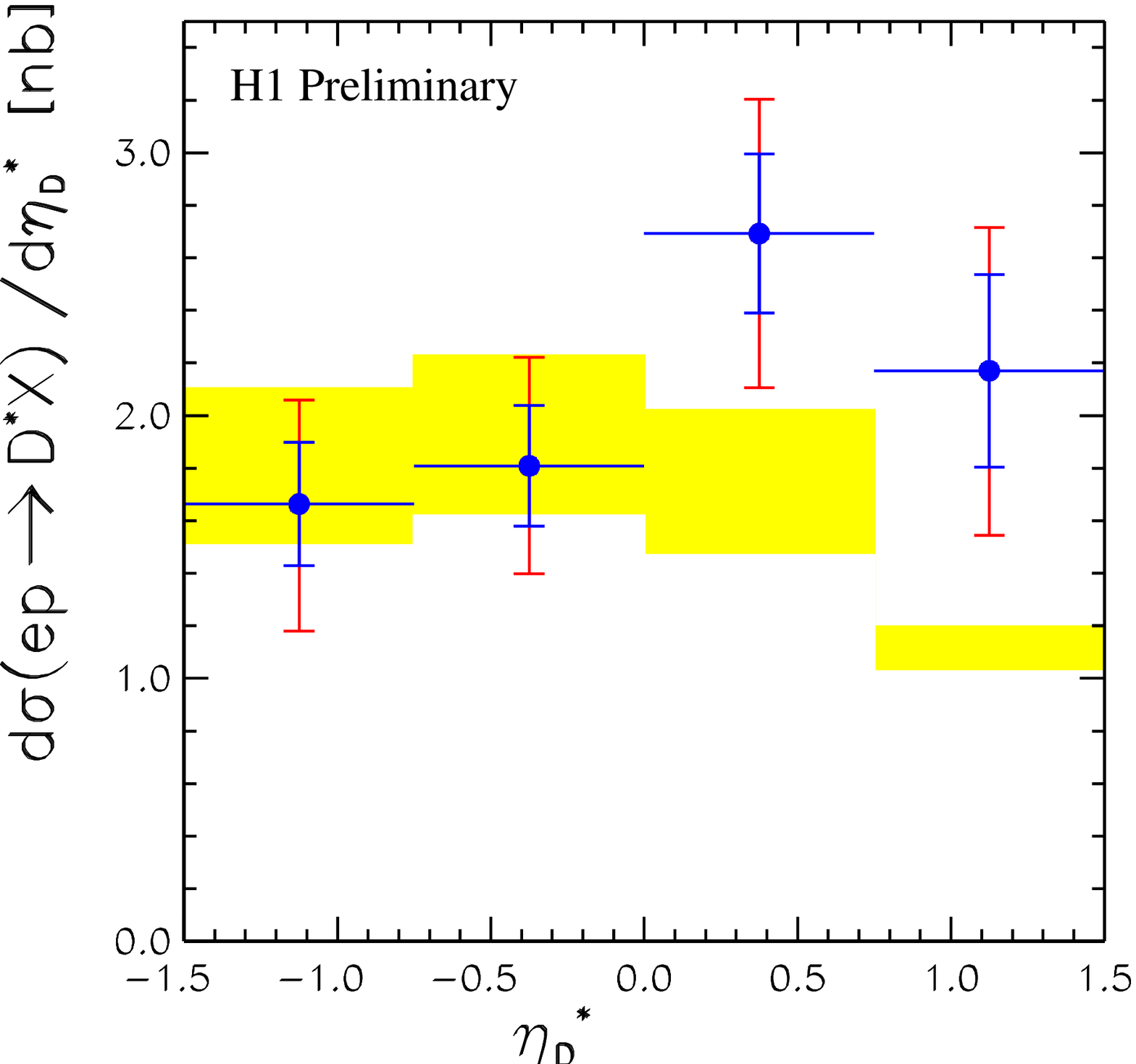,height=5.1cm,clip=}
\epsfig{file=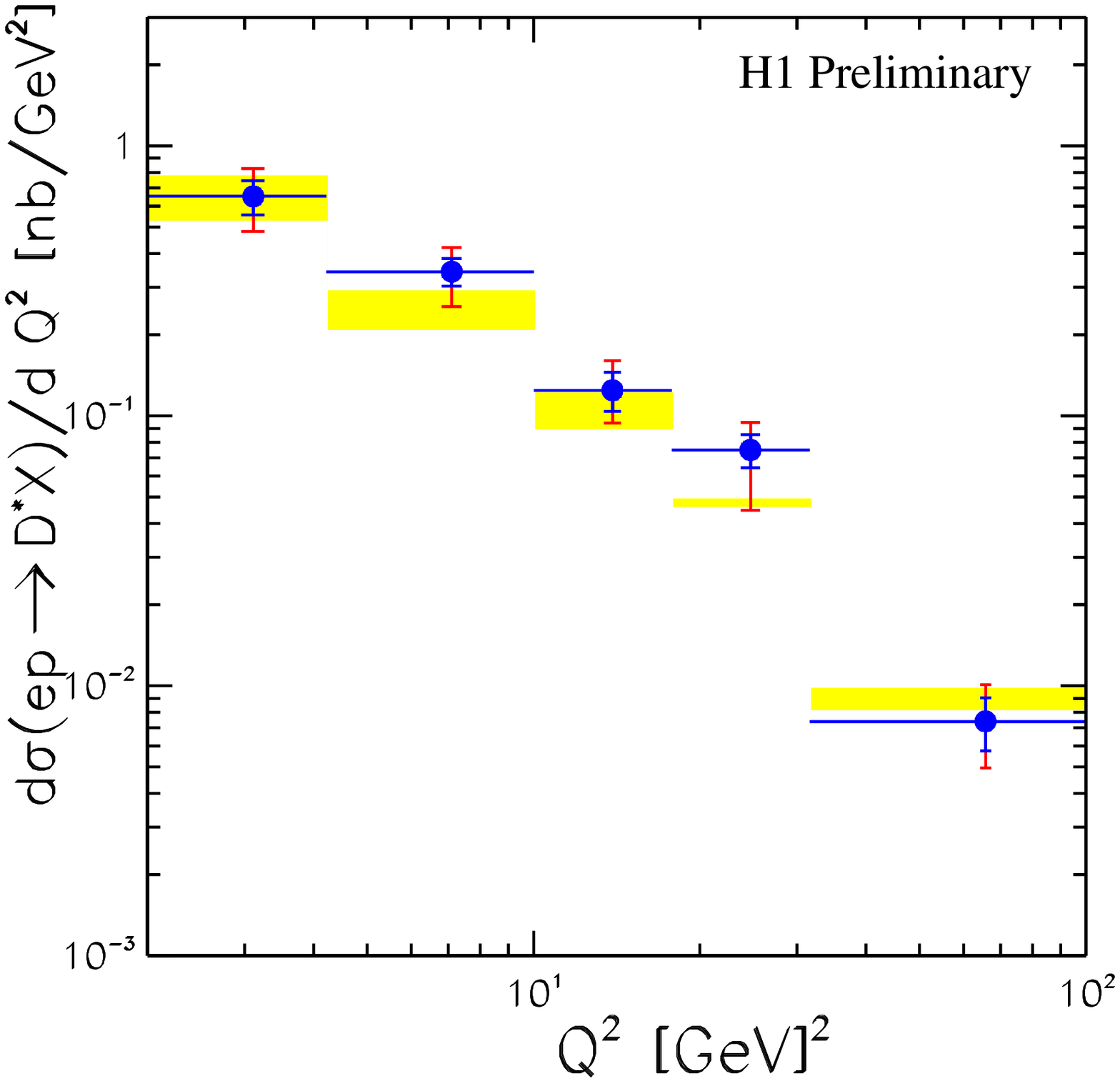,height=5.1cm,clip=}
\caption {H1: Cross sections ($2<Q^2<100\,$GeV$^2$, $0.05<y<0.7$,
$p_T^{\protect\dstar}>1.5\,$GeV, $|\eta^{\protect\dstar}|<1.5$) compared to NLO 
calculations~\protect\cite{harris}
. The shaded band represents the uncertainty in $m_c$.}
\label{fig3}
\end{figure}

\begin{figure}[t!]
\epsfig{file=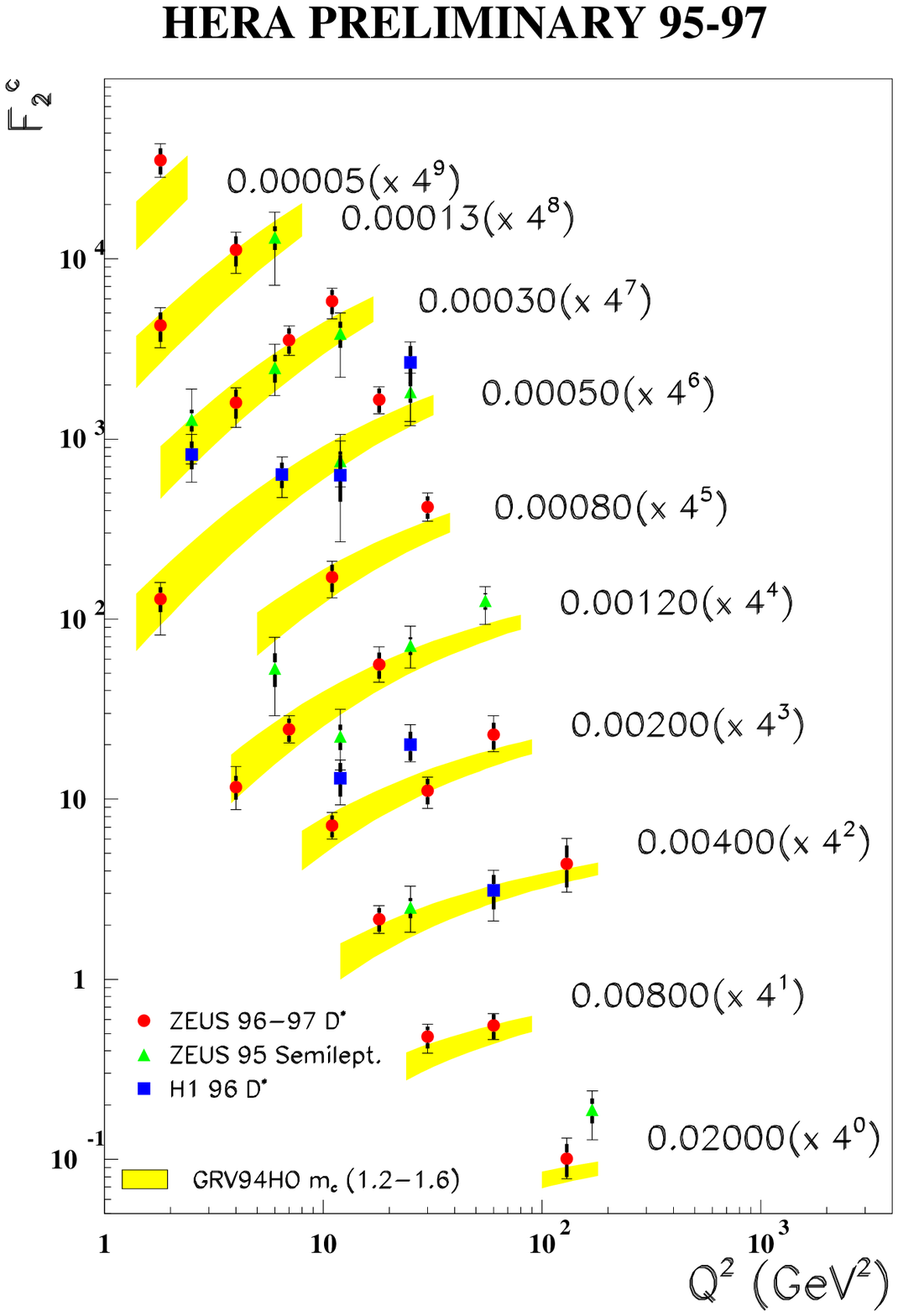,width=7.0cm,
                                bbllx=0pt,bblly=0pt,bburx=390pt,bbury=560pt}
\epsfig{file=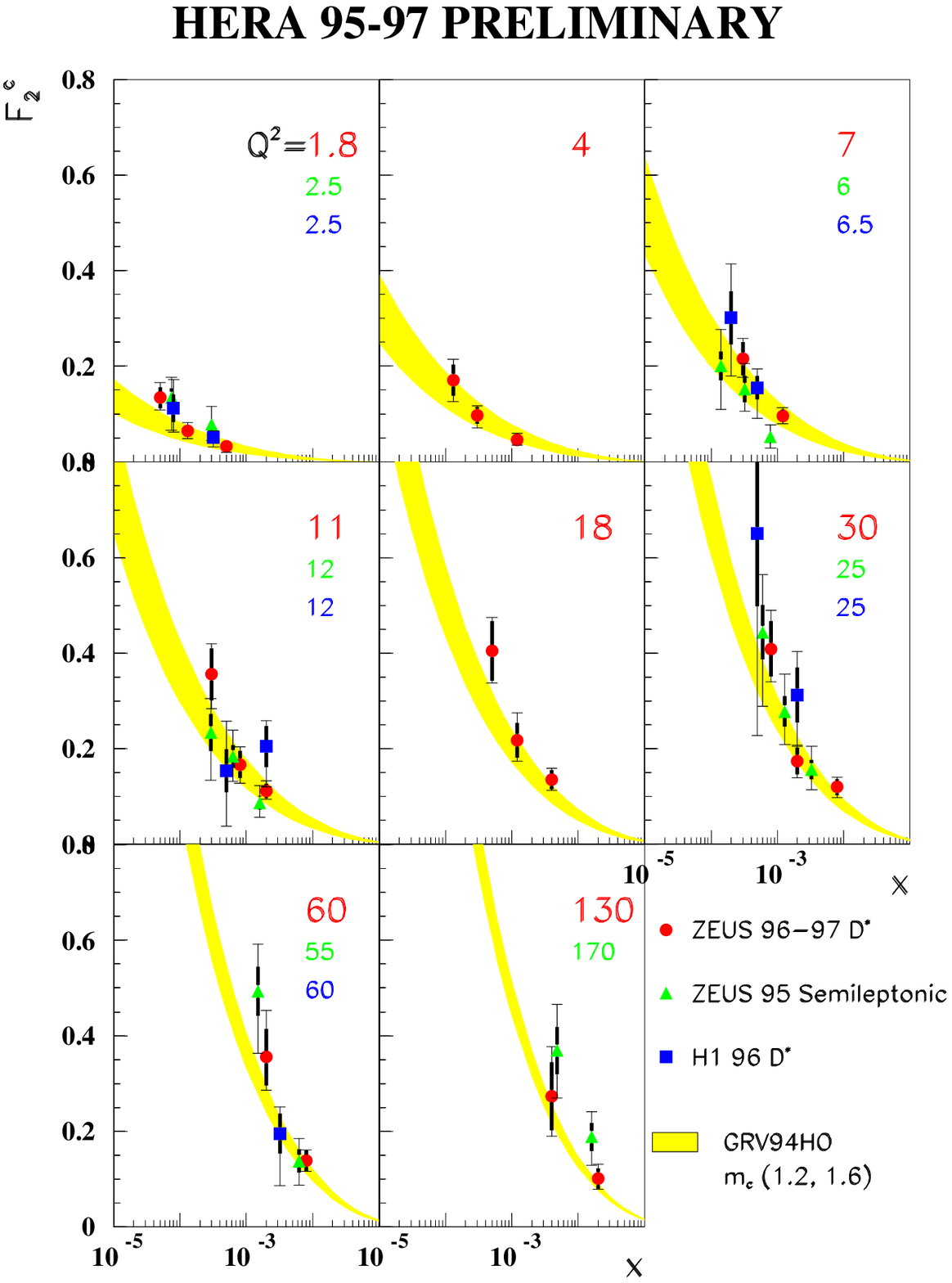,width=7.2cm,
                             bbllx=0pt,bblly=0pt,bburx=400pt,bbury=565pt}
\caption {\protect$\ftcc$\ as function of \protect\qsq\ at fixed $x$ (scale factors applied for clarity); \protect$\ftcc$\ as function of $x$ in bins of 
\protect\qsq. For comparison a NLO prediction using the GRV94HO parton densities is shown with the uncertainty due to the charm mass.}
\label{fig4}
\end{figure}

The integrated luminosity delivered by  HERA has steadily increased over the 
years. This review
will cover data from 1995 ($\sim 6\,\picb$),\ 1996\ ($\sim 10\,\picb$)
and 1997 ($\sim 26\,\picb$).
Most results are preliminary.

\medskip
The usual kinematic variables for deep inelastic scattering are used:

$s=(k + P)^2;\, Q^2=- (k - k')^2;\,x=  \frac{Q^2}{2 P \cdot q};\,
y=\frac{q \cdot P}{k \cdot P};\,
W^2_{\gamma p}  =  (q + P)^2 = s y -\qsqw$

where $k$ and $P$ are the four vectors of incoming electron and proton, 
and $q$ of the exchanged photon.

\subsection*{Determination of {\protect\boldmath$F^c_2$}}

The inclusive cross section for production of charm in deep inelastic scattering (DIS) can be written as

\begin{displaymath}
\frac{d^2 \sigma^{ep\ra e\,c\bar{c}\,X}}{dx dQ^2} = \frac{2 \pi \alpha^2 }{x
Q^4}
(1+(1-y)^2)\cdot F^c_2(x,Q^2)
\end{displaymath}

where the contribution due to $F_L$ has been neglected since it is
expected to be small. 

Charm is tagged through reconstruction of $D^{*+} \rightarrow D^0 \pi^+$ 
with subsequent decay $D^0 \rightarrow K^- \pi^+$ and also the 
charge conjugate decay.
ZEUS has presented a new analysis of semileptonic
charm decays $c \rightarrow e^+ + X$. The electron was identified
using the electromagnetic calorimeter and the specific energy loss $dE/dx$ in
the driftchamber. Details of the analysis from H1 and ZEUS 
can be found in~[1b].

For $D^*$ production a comparison to the RAPGAP 
Monte Carlo~\cite{rapgap} simulation
is shown in fig.~\ref{fig2}. Reasonable agreement is found. After
unfolding detector effects cross sections are obtained in a restricted 
kinematical region, 
examples from H1 data are shown in fig.~\ref{fig3}. The data
are compared to a NLO calculation by Harris and Smith~\cite{harris} using 
the Peterson
fragmentation function. The agreement is good and the extrapolation to the
full kinematic region is done with this calculation. 

The resulting $Q^2$ and
$x$ dependence of $F^c_2$ is shown in fig.~\ref{fig4}.
The data span $Q^2$ values from 1.8 to 130 GeV$^2$ and $5 \cdot 10^{-5} \leq
x \leq 0.02$.
The agreement of the different data sets is reasonable within errors.
Also shown is the theoretical NLO calculation using the GRV94-HO parton
density functions which reproduces the data well. A strong rise of $F^c_2$
towards low $x$ is observed at fixed $Q^2$ and in $Q^2$ strong scaling
violations are seen at fixed $x$.
$F^c_2$ gives a contribution of between 10\% (low $Q^2$) and 30\% (high $Q^2$)
to the inclusive $F_2$ at an $x \sim 5 \cdot 10^{-4}$.

\begin{figure}[pt!]\centering
\epsfig{file=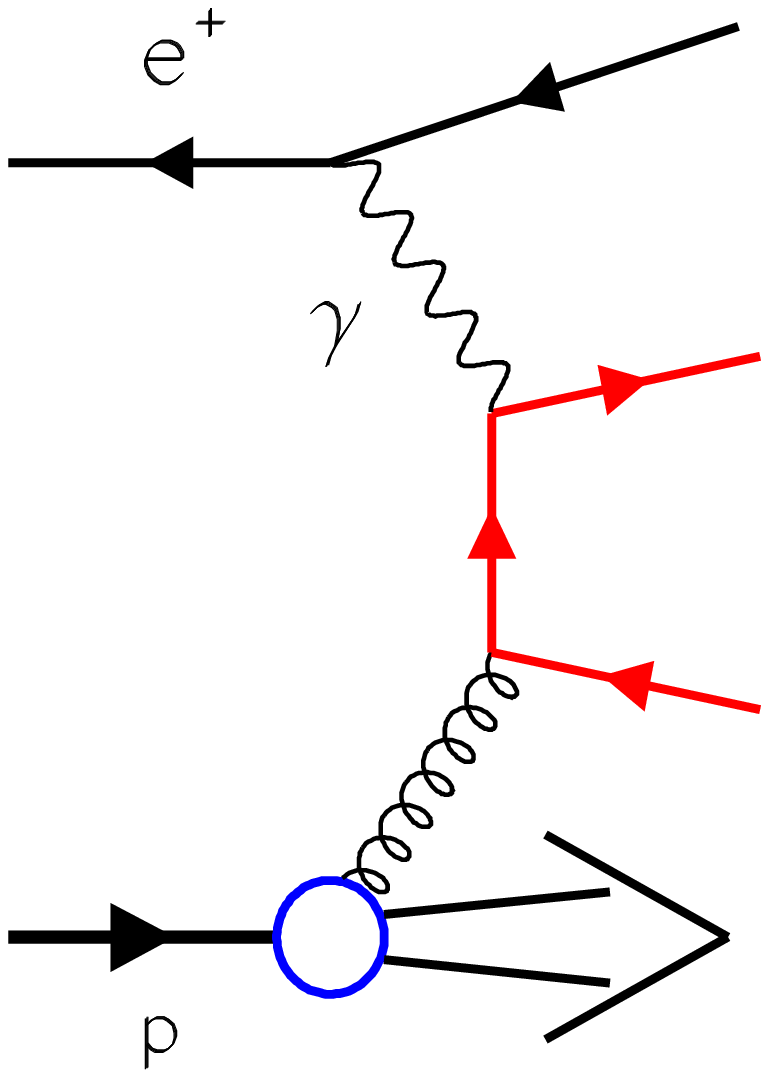,width=2.5cm,bbllx=100pt,bblly=120pt,bburx=360pt,
                   bbury=440pt,clip=}\qquad
\epsfig{file=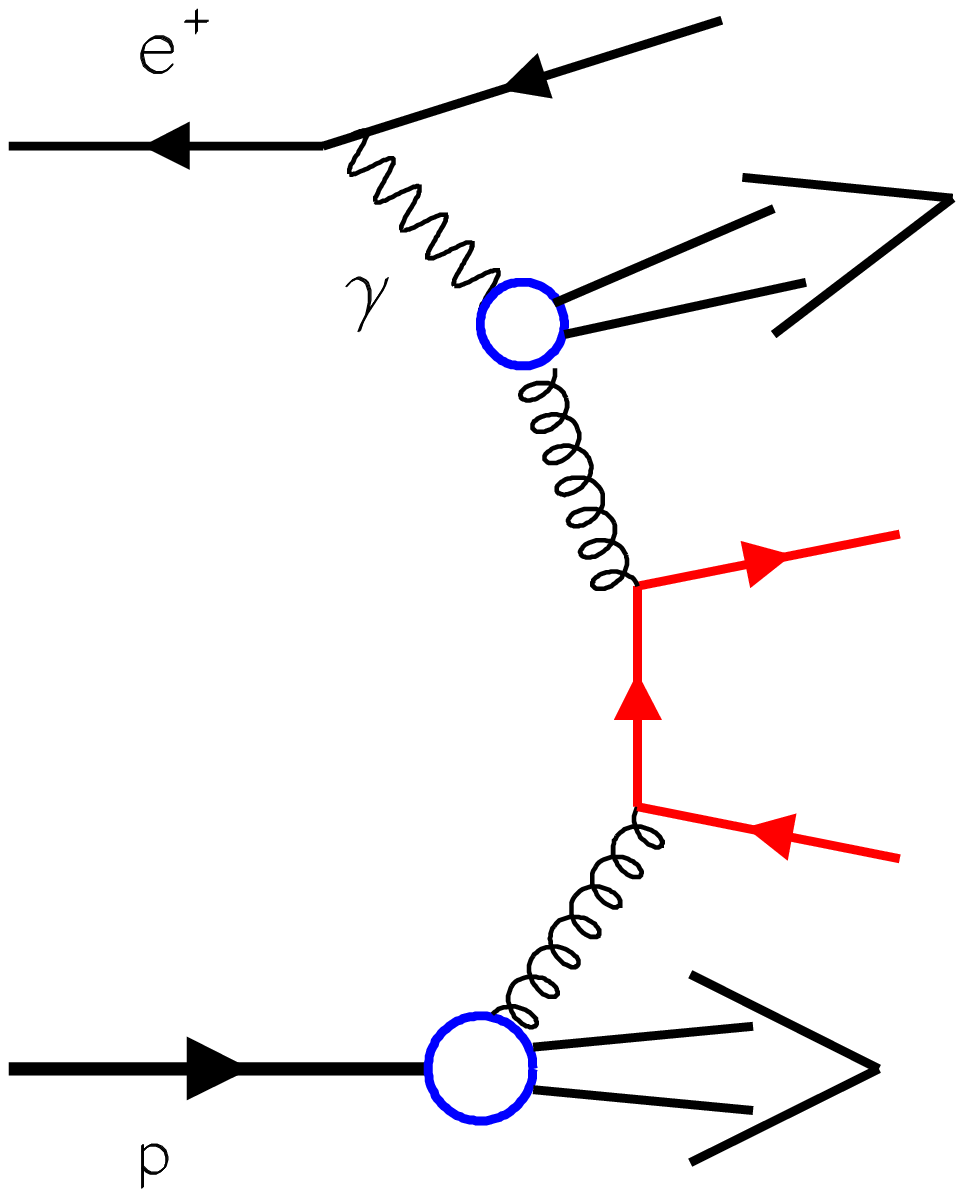,width=2.5cm,bbllx=100pt,bblly=120pt,bburx=360pt,
                   bbury=440pt,clip=}\qquad
\epsfig{file=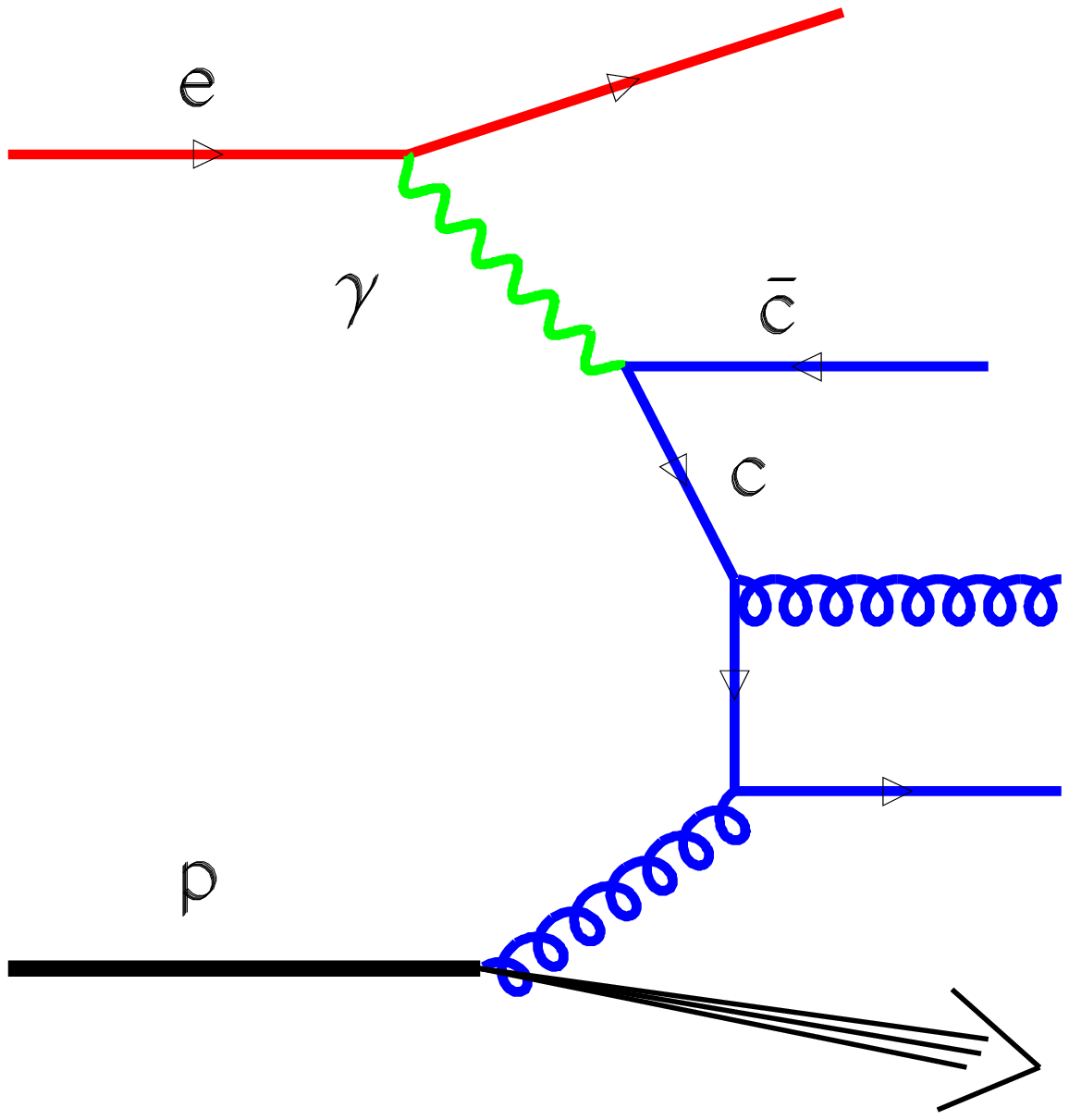,width=3.2cm,bbllx=40pt,bblly=130pt,bburx=430pt,
                   bbury=500pt,clip=}
\caption{Generic diagrams for a) direct, b) resolved charm production. In c)
a NLO diagram is shown (flavour excitation).}
\label{fig5}

\epsfig{file=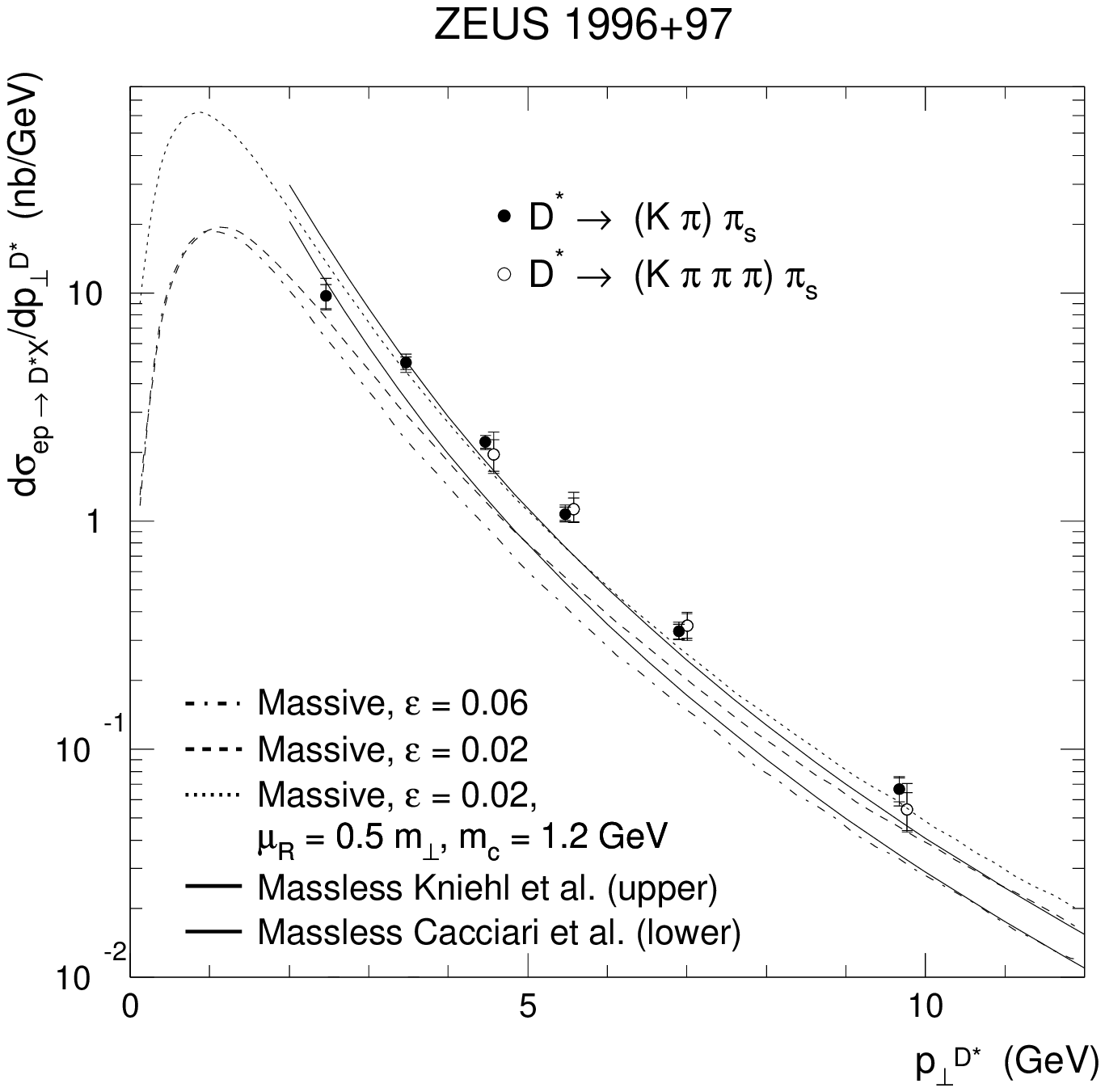,width=8.5cm,
                                bbllx=50pt,bblly=20pt,bburx=400pt,bbury=400pt}
\caption{$d\sigma/dp_\perp$ for  $D^*$ photoproduction from 
ZEUS for $130<W_{\gamma p}<280\,\mbox{GeV}$,
$Q^2 < 1\ \mbox{GeV}^2$ and $|\eta^{D^*}|<1.5$\  
($\eta^{D^*}$ is the pseudorapidity of the $D^*$) 
for the $(K\pi)\pi$ and
$(K\pi\pi\pi)\pi$ channels. The curves represent ``massless'' and ``massive'' calculations as indicated.} 
\label{fig6a}

\setlength{\unitlength}{1cm}
\begin{picture}(16.0,4.8)
\put(3,0){\epsfig{file=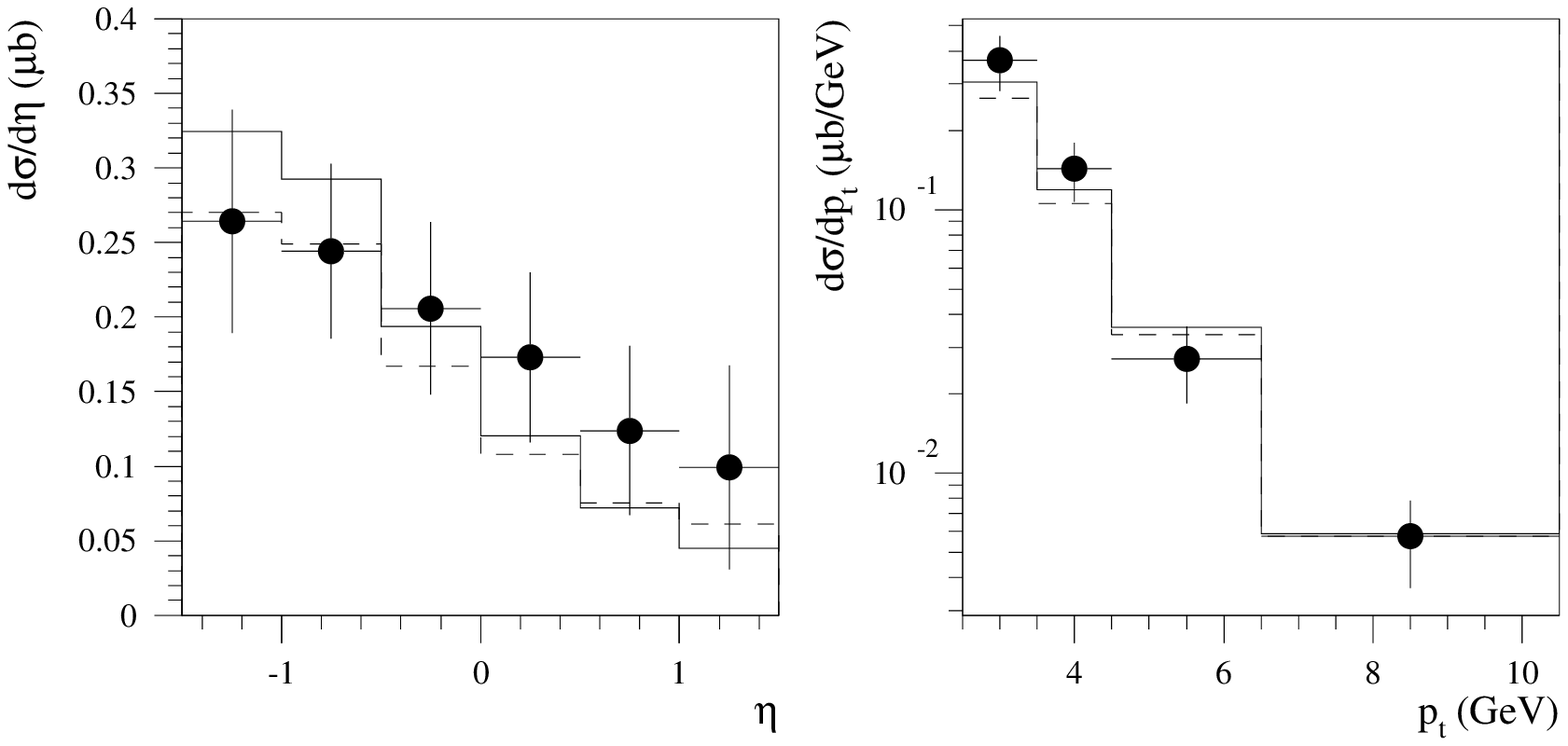,height=4.5cm,
                        bbllx=30pt,bblly=285pt,bburx=275pt,bbury=545pt,clip=}}
\end{picture}
\caption{Photoproduction of $D^*$ from H1: 
$d\sigma^{\gamma p}/d\eta$ for $p_\perp>2.5\,$GeV
and $d\sigma^{\gamma p}/dp_\perp$ for $|y^{D^*}|<1.5$, where $y^{D^*}$ 
is the rapidity.
The histograms show the NLO QCD predictions calculated according to 
\protect\cite{frixione} with the MRSG (solid) and MRSA' (dashed)  
proton parton density parametrizations.}
\label{fig6b}
\end{figure}

\subsection*{Photoproduction of {\protect\boldmath$D^*$}}

When the exchanged photon is almost real, contributions due to its hadronic
nature have to be taken into account (``resolved'' processes). 
In NLO QCD calculations an unambiguous separation
of the direct process (fig.~\ref{fig5}a) and resolved processes (b and c)
is no longer possible, only the sum of the two is well defined.
There are two approaches to calculate the photoproduction cross %
sections in \nlo:

{\em The ``massive'' approach}~\cite{frixione} where only the light quarks 
u, d and s and gluons are active partons
in the photon (and proton). Charm is only generated in the hard subprocess 
(see also fig.~\ref{fig5}b).
This approach is valid for $m_c \gg  \Lambda_{QCD}$.
{\em The ``massless'' approach}~\cite{kniehl,cacciari} where also 
charm is an active flavour. This approach
is valid at $p_t \gg m_c$.     

The high statistics data from ZEUS \cite{dstarz-vanc} 
are shown in fig.~\ref{fig6a}. They are found to be above
the massive and massless calulations.
The comparison of H1 data~[1b] with massive calculations shown 
in fig.~\ref{fig6b}
is satisfactory. 

ZEUS has presented an analysis of $D^*$ events  which
contain  two jets~\cite{dstarz-vanc}.
In these events the {\em observed} momentum fraction $x^{obs}_\gamma$ can
be calculated, which 
describes the fraction of the photon energy contributing to the production 
of the two jets.
A significant tail at low $x^{obs}_\gamma$ is found in the data.  
In the generator HERWIG this tail can be 
described by charm excitation in the photon,
while considering only light flavours leads to  
discrepancies.


\begin{figure}[t!] 
$\begin{array}{ll}
  \begin{minipage}{.38\linewidth}
  \epsfig{file=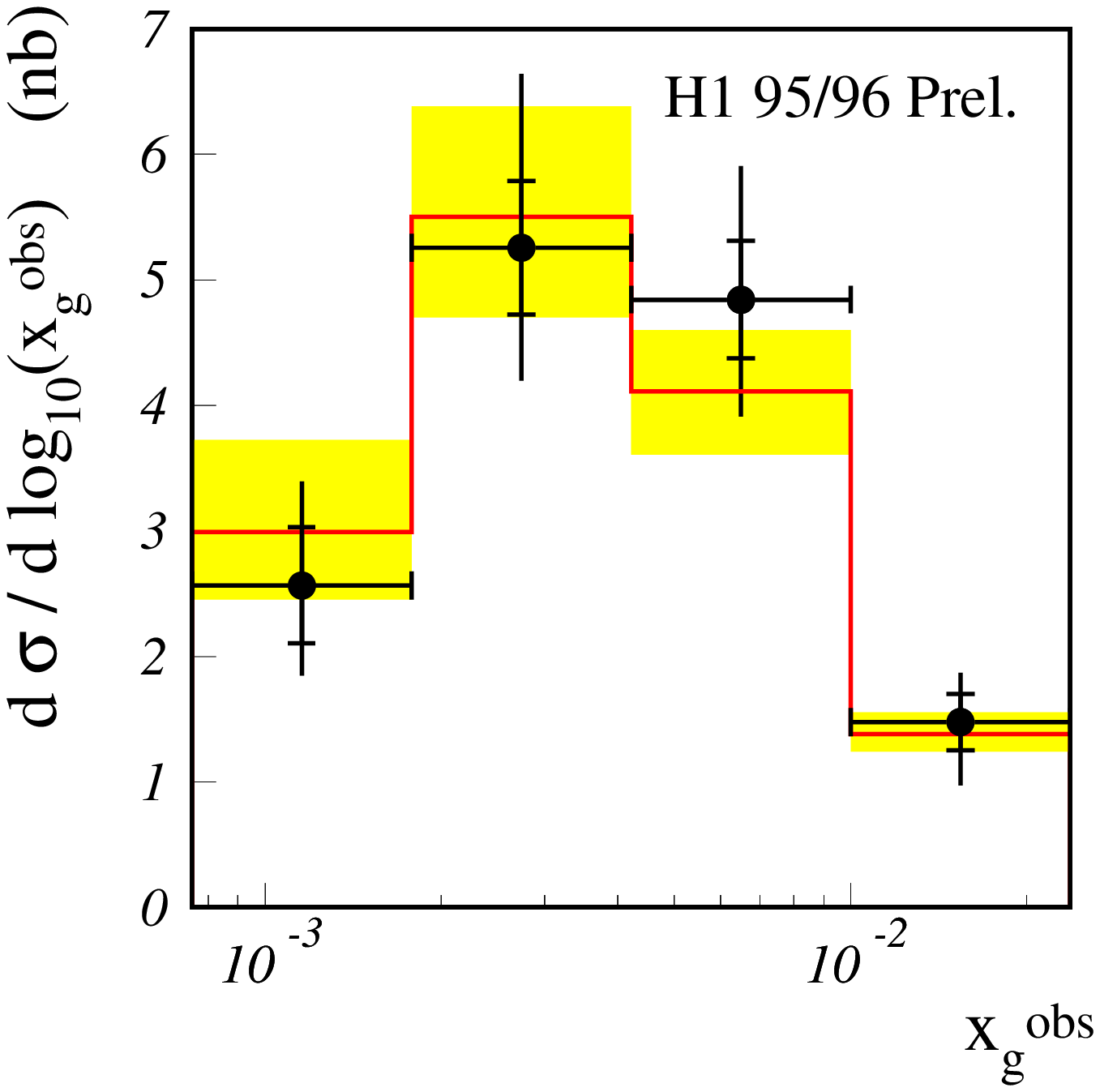,
    bbllx=30pt,bblly=10pt,bburx=410pt,bbury=410pt,width=5.4cm,clip=}
  \epsfig{file=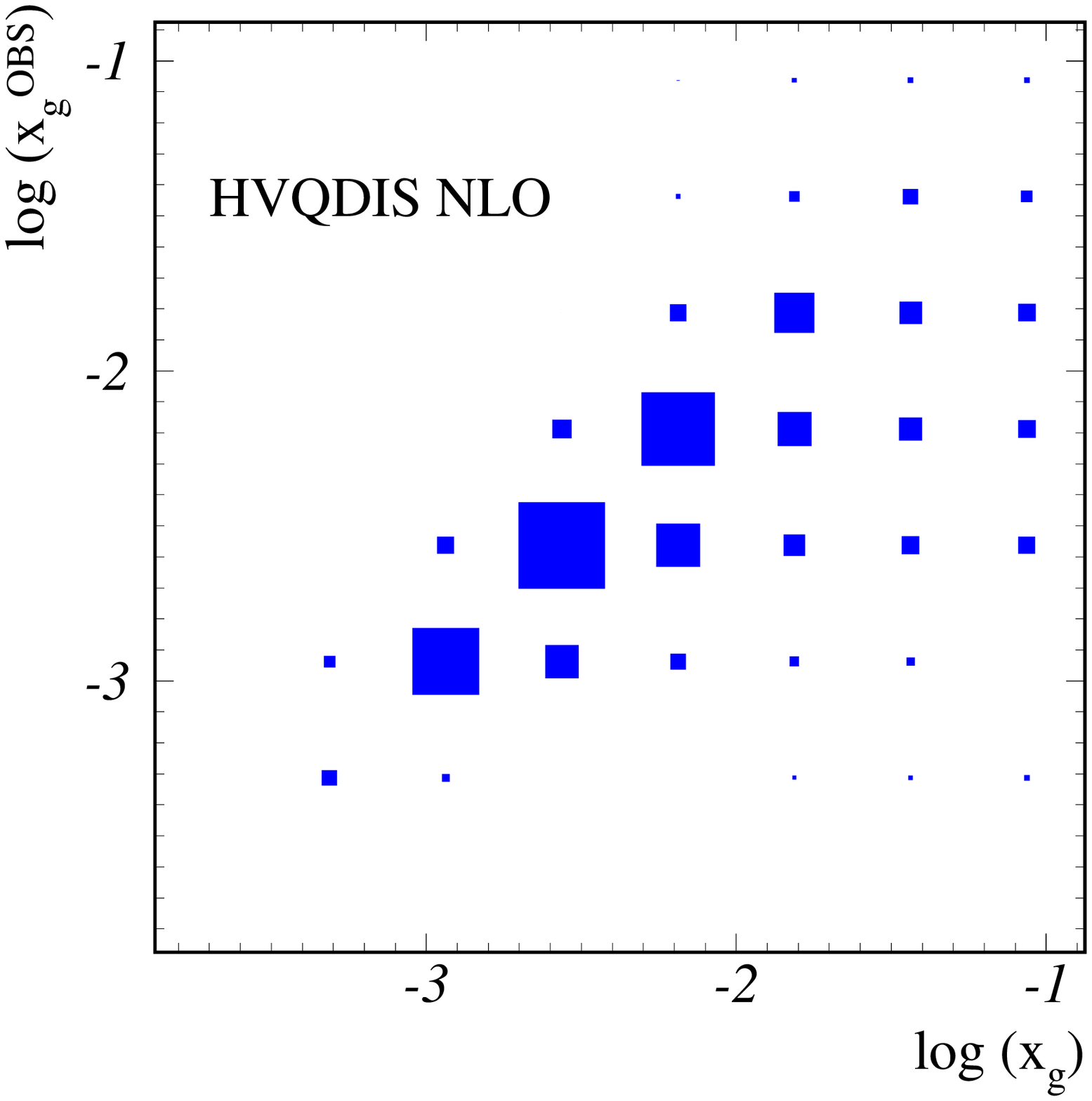,
          bbllx=10pt,bblly=0pt,bburx=530pt,bbury=530pt,width=5.5cm,clip=}
   \end{minipage}&
  \begin{minipage}{.56\linewidth}
          \epsfig{file=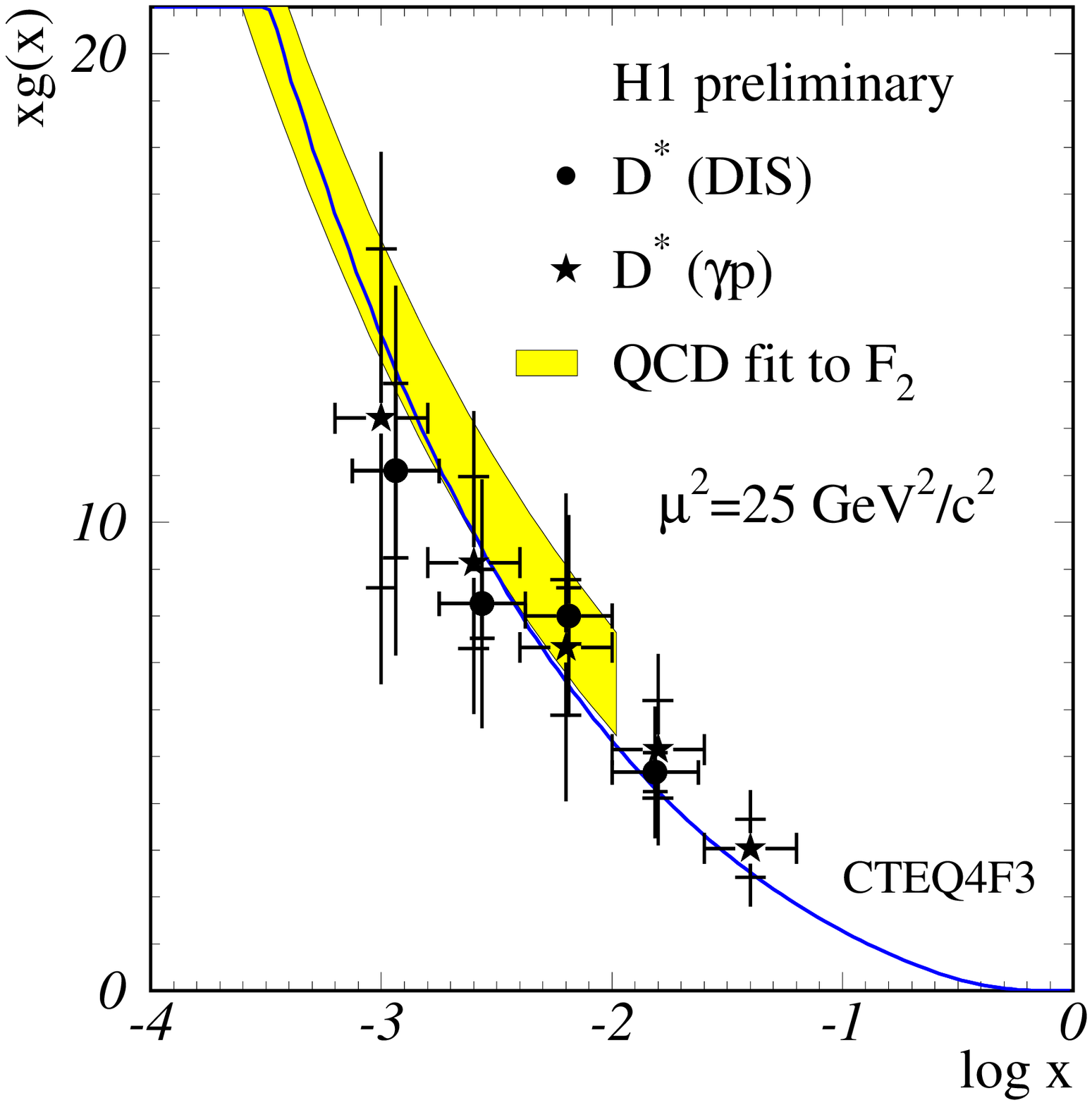,
              bbllx=60pt,bblly=0pt,bburx=560pt,bbury=560pt,width=9.1cm,clip=}
 \end{minipage}
\end{array}
$
\caption{(a) Differential cross section for 
$ep \rightarrow eD^*X$ as 
function of $x_g^{obs}$ in the visible range compared to 
 the NLO QCD prediction using the CTEQ4F3 parton
distribution. The shaded band reflects the uncertainty due to 
the charm mass $1.3<m_c<1.7\,$GeV. 
(b) Correlation between $x_g^{obs}$
and true momentum fraction $x_g$.
(c) Comparison of the gluon densities obtained from the two 
$D^*$ analyses in DIS and photoproduction.
The systematic error is dominated
by the uncertainty of the charm quark mass and the fragmentation parameter.
For comparison the H1 QCD analysis of the inclusive
$F_2$ measurement (shaded band)
and the CTEQ4F3 parametrization are shown.}
\label{fig7}
\end{figure}

\subsection*{Gluon density from {\protect\boldmath$D^*$ events}}

H1 extracted the proton's gluon density function in DIS ($2<Q^2<100\,\mbox{GeV}^2,\,0.05<y<0.7$) 
and in photoproduction ($0.02<y<0.32;\,0.29<y<0.62;\,Q^2\sim 0$)~[1b].

The observed momentum fraction $x_g^{obs}$ of the gluon is reconstructed
 from the kinematics of the final state and a differential
cross section $d\sigma/dx_g^{obs}$ is determined, which for the DIS data
is shown in fig.~\ref{fig7}. The correlation of $x_g^{obs}$  with the
true $x_g$ as given by the NLO QCD calculations~\cite{harris} -- also
shown in fig.~\ref{fig7} -- is used in an iterative
unfolding procedure to obtain $d\sigma/dx_g$. The gluon density is then
obtained by reweighting the calculation  with the measured
cross section. The result is shown in fig.~\ref{fig7} as a momentum
distribution $x_g \cdot g(x_g)$. The range $10^{-3} < x_g < 0.02$ is covered.
The data from photoproduction and DIS agree well within the large errors.
They also agree with the result from an analysis of scaling violations
in the inclusive measurement of $F_2$~\cite{h1newglue}.


\subsection*{\protect\boldmath$b\bar{b}$ Production}

Due to the higher mass of the $b$ quark the total \cs\ for 
$\bbbar$\ production is
expected to be 200 times smaller than that for $\ccbar$\ production.
The theoretical uncertainties in calculating the \nlo\ predictions 
are, however, smaller~\cite{btheor}. H1 determined the
\cs\ for the first time in the HERA energy range using semi--muonic $b$ decays~[1c].

A photoproduction event sample was selected containing
two jets of transverse energy $E_T >$ 6 GeV 
and a muon of transverse momentum (relative to the beam direction) 
$p_T^\mu >$ 2 GeV in the central detector 
region $35\degree < \theta^{\mu} < 130\degree$.

\begin{figure}[t!]
\epsfig{file=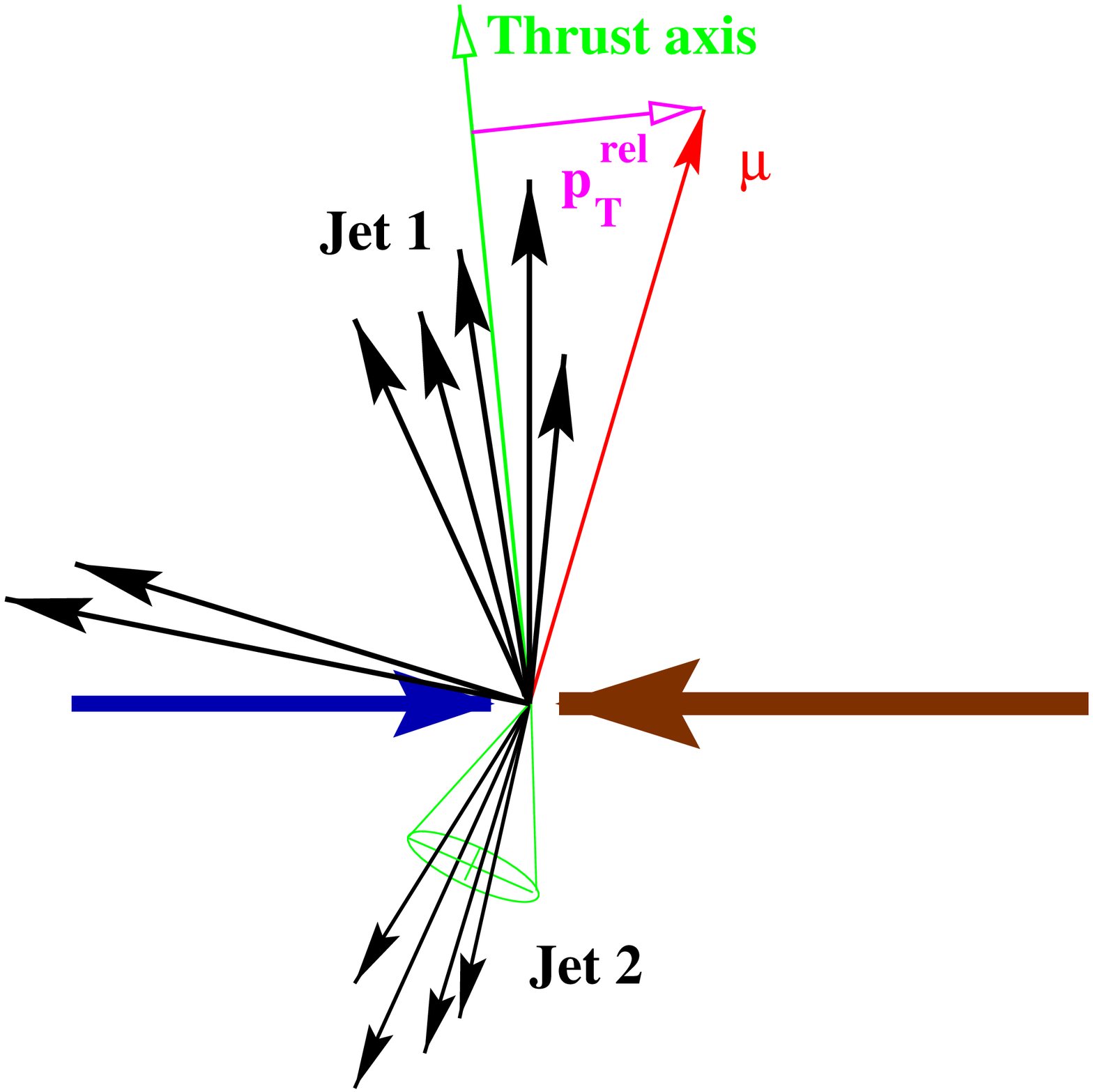,width=6.5cm,clip=}\hspace{-0.5cm}
   \epsfig{file=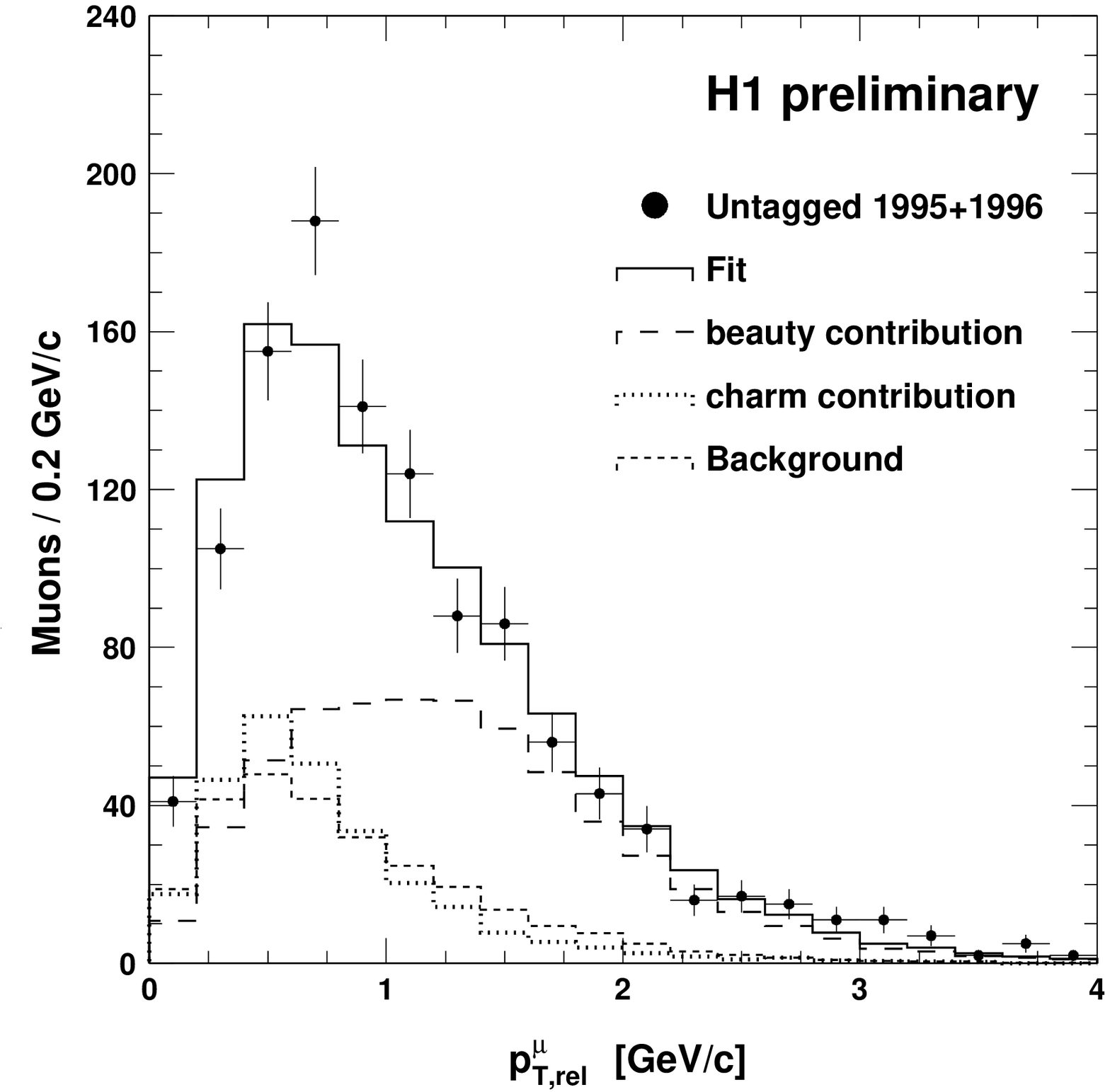,height=7.9cm,clip=}
\caption{{\em Left:} Definition of $p^\mu_{T,rel}$. {\em Right:} The measured $p^\mu_{T,rel}$ 
distributions and the result of the fit
(solid line); the contributions from beauty, charm
and background are shown separately.}
\label{fig8}
\end{figure}

The thrust axis\footnote{The thrust axis is the axis which maximizes
$T = \mbox{max} (\frac{\Sigma p_i^L}{\Sigma |p_i|})$, where the sum runs over
all particles belonging to the jet and $p_i^L$ is the
component of the particle momentum parallel to the thrust axis.}
was determined for each jet in order to approximate the $b$ flight
direction. The transverse momentum of the muon $p^\mu_{t,rel}$ with respect
to the jet is used as a discriminating variable: Muons from $b$ decay
show a $p^\mu_{t,rel}$ spectrum extending to higher values than $c$
decays (see fig.~\ref{fig8} for an illustration of the method).

The background comes from the production of the light quarks $u, d$ and
$s$, which is roughly a factor 2000 larger than $b$ production. Punch
through and decay in flight lead to false muon signatures. The contribution
is determined from data using an independent dataset and using the muon
fake probability and the hadron composition from a well tuned and checked
simulation program.
The resulting $p_{t,rel}^\mu$ spectrum is shown in fig.~\ref{fig8}
indicating the background contribution (23.5\%), which is absolutely
determined. The fractions of $b$ and $c$ quarks are obtained from a fit to 
the data distribution, yielding
(51.4 $\pm$ 4.4)\% and $(23.5 \pm 4.3)$\%, respectively. 

The cross section in the visible kinematic range of
$Q^2 < 1\,\mbox{GeV}^2;\,p_T^\mu > 2\,\mbox{GeV}$;
$95\leq W_{\gamma p}\leq 270\,\mbox{GeV};\,35\degree<\theta^\mu < 130\degree$
is determined as

$$\sigma(ep \rightarrow e\,\bbbar + X)^{vis} = \,0.93 \pm 0.08^{+0.21}_{-0.12}
\,\mbox{nb},$$

where the first error is statistical and the second systematic.
Contributions to the systematic errors are  
the branching ratio $b\ra X\mu\nu$, the energy scale of 
calorimeters and detector efficiencies. 

The corresponding direct LO cross section from the AROMA~\cite{AROMA} 
simulation is $0.19\,\mbox{nb}$, roughly a factor 5 lower.
The fraction of $c$ quarks determined from this analysis leads to the same
\cs\ for $ep \rightarrow e\ccbar X$ as previously determined
from analysis of $D^*$ production~\cite{ccbar}.

\subsection*{Inelastic {\protect\boldmath$J/\psi$ production}}
\normalsize
New data on charmonium ($J/\psi,\ \psi'$) and $\Upsilon$ production  have been
presented by H1 and ZEUS~[1a]. Here we will concentrate on
``inelastic'' $J/\psi$ production as opposed to the diffractive processes
which dominate the \cs\ at low $Q^2$. Inelastic 
$J/\psi$ production
could at lower $W_{\gamma p}$ (fixed target regime) be well described by
the Colour Singlet Model (CSM).
For HERA the CSM \cs\  calculations are available in NLO~\cite{Kraemer}.

\begin{figure}[t!]
\epsfig{file=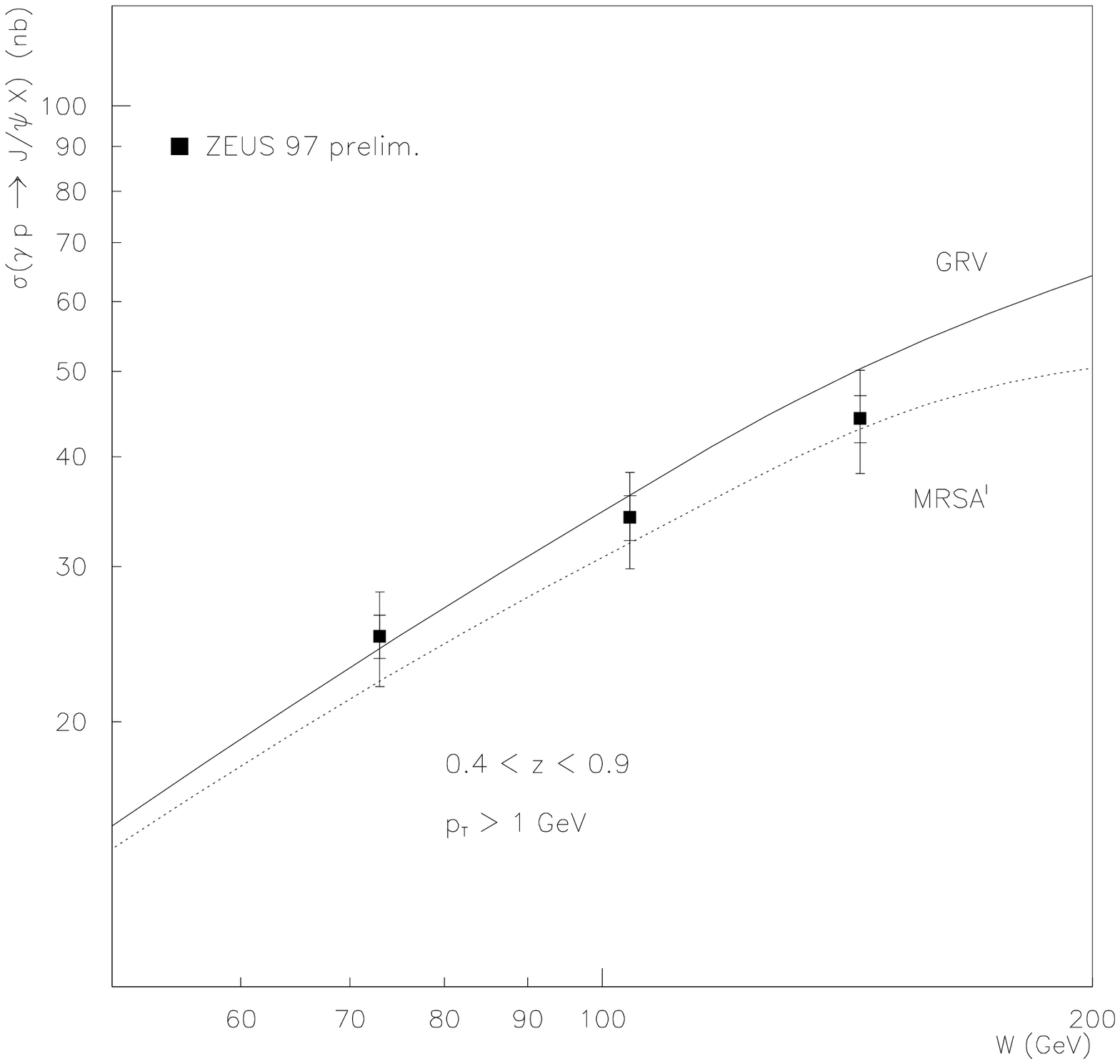,height=7.3cm,clip=}
\epsfig{file=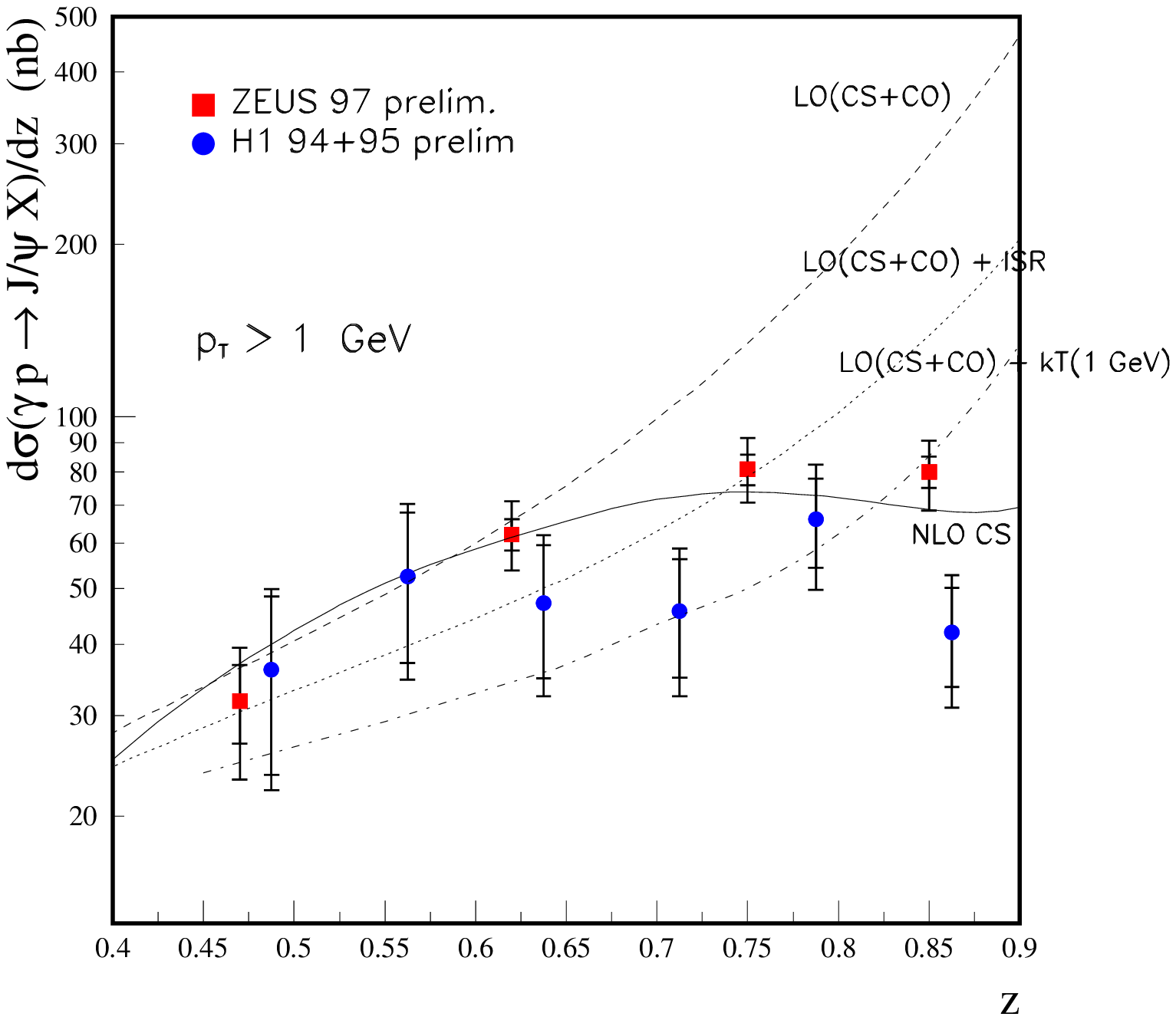,height=7.3cm,clip=}
\caption{Inelastic $J/\psi$ photoproduction: a) cross section as 
function of $W_{\gamma p}$ for $0.4 < z < 0.9$ and $p_T > 1\,\mbox{GeV}$. 
The lines
correspond to the NLO predictions from \protect\cite{Kraemer}
with GRV or MRSA' [20] parton density functions
($m_c = 1.4$ GeV and $\Lambda_{QCD} = 300\ \mbox{MeV}$).
b) $d\sigma/dz$ for $50<W_{\gamma p}<180$\,GeV and $p_T>1$\,GeV. 
The NLO computation 
is shown as a solid line. The dashed and dotted
lines are given by the sum of the colour--singlet and the colour--octet
leading order calculations performed in~\protect\cite{Cano} and 
in~\protect\cite{Martin}. The latter
include estimates of higher order QCD corrections due to initial state 
radiation.}
\label{fig10}
\end{figure}

As is well known the CSM fails to describe charmonium production in
$p\bar{p}$ collisions at high $p_T$~\cite{fail}. Colour octet
contributions have been proposed for an adequate description. The NRQCD 
factorisation approach (NRQCD = Non Relativistic QCD)
describes any process $A + B \rightarrow J/\psi + X$ as a sum over colour
singlet and colour octet contributions.


Whereas the transformation of a colour singlet $^3S_1$ state into
a $J/\psi$ can be calculated using the measured semileptonic decay width, the
transition of a colour octet state to $J/\psi$ is non--perturbative and
at present not calculable.
Therefore predictions for the \cs\  at HERA use the
non perturbative transition matrix elements extracted from the CDF data.

\begin{figure}[t!]\centering
\epsfig{file=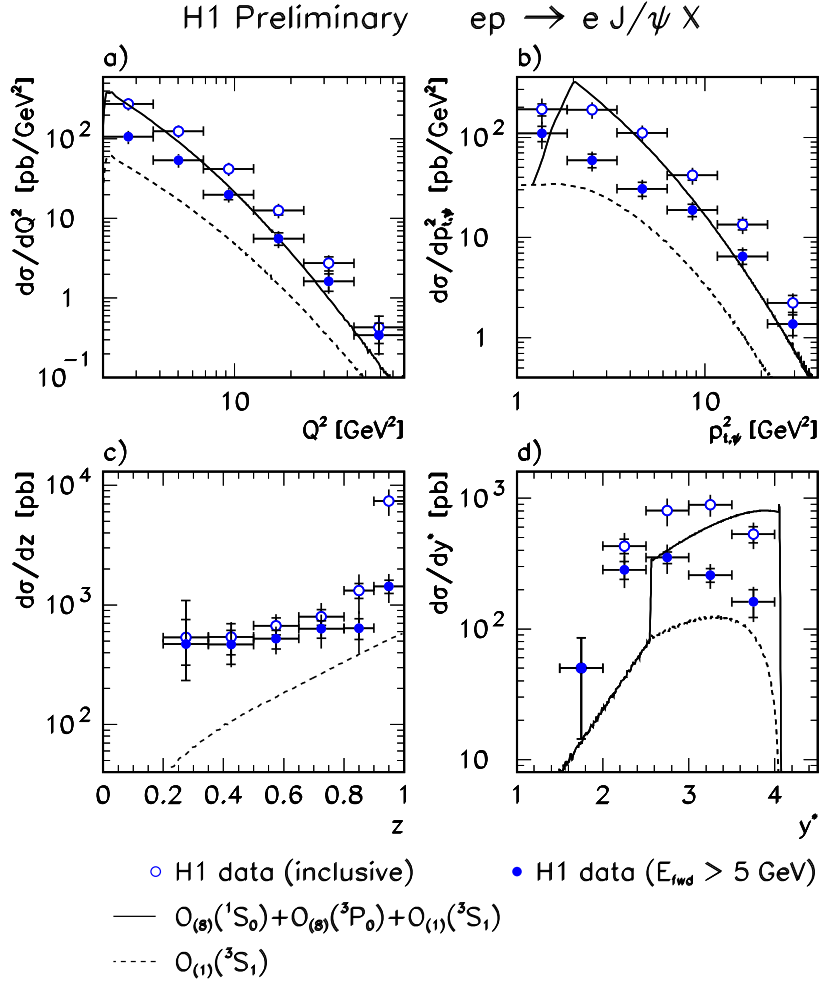,height=15cm,clip=}

\vspace{-1cm}
\caption{Differential cross sections for the inclusive 
                (open points) and inelastic ($E_{fwd}>5\:$GeV, black points)
                     $ep \rightarrow e\:J/\psi\: X$ process.
        a) $d\sigma / dQ^2$, b) $d\sigma / dp_{t,\psi}^2$, c)
        $d\sigma / dz$ and d) $d\sigma / dy^\ast$. 
        The curves are predictions \protect\cite{fleming} within
        the NRQCD factorization approach for the colour singlet contribution 
        (dashed) and the
        sum of singlet and octet contributions (full line).}
\label{fig11}
\end{figure}

The ZEUS collaboration
has updated their photoproduction data~[1d]. 
 The results for the $\gamma p$
\cs\  as function of $W_{\gamma p}$ and of $z$ are shown in
fig.~\ref{fig10}. The data agree well with the
\nlo\ pQCD calculation in the \colsingmodel~\cite{Kraemer}.
The variable $z$ is defined as $z = \frac{P_\psi \cdot P}
{P \cdot q} \approx \frac{E_\psi}{E_\gamma}$ where the latter 
approximation holds in the
proton rest frame. 
In fig.~\ref{fig10} b in addition to the CSM in NLO calulations using
the NRQCD/factorisation approach~\cite{com,Cano,Martin} are shown. 
The upper curve was calculated
in LO using the transition matrix elements extracted from CDF data in LO 
and shows a strong rise towards high $z$ values. The lower curves
also take into account higher orders {\em approximately}  
as explained in refs.~\cite{com,Cano,Martin}. 
Doing so leads to modifications in the non perturbative matrix elements 
and/or in the \cs s themselves. 
The net effect is a decrease of the predicted rise at high $z$.

H1 has for the first time determined the \cs s for inelastic
$J/\psi$ production at $Q^2> 2\ \mbox{GeV}^2$~[1d]. 
The results are shown
in fig.~\ref{fig11}. Two data sets are shown, a completely inclusive one
(open points) and one where the diffractive contributions have been
removed by a cut on the energy in the forward region of the detector as
suggested by Fleming and Mehen~\cite{fleming}, 
whose LO calculations are shown for comparison.
The data are seen to be far above the CS contributions. The magnitude
of the data is reproduced better taking into account colour octet
contributions. The shape of the latter leaves, however, much room for
improvement, in particular in the rapidity $y^*$ in the $\gamma^*\,p$ 
\cms\ system. Note that the NRQCD calculations are performed at the
parton level, no smearing due to the transition into $J/\psi$ is
taken into account.

\subsection*{Summary}
Due to increased statistics detailed analyses of heavy flavour production in
$ep$ collisions are performed in a variety of channels and kinematic regions.
$\bbbar$\ production was observed for the first time in photoproduction
via semi--muonic decay of  the b--quark. The \cs\ was found to be 
considerably larger than the leading order predictions for the direct process.\\
In the range $2\lsim$\qsq$\lsim130$\,\gevsq \cs s and the charm contribution 
to $\ftcc$\ are determined and 
found to agree with \nlo\ predictions.
In photoproduction the validity of different approaches to calculate
\nlo\ corrections is being studied in various kinematic regions.

Since photon gluon fusion is the dominant process a direct 
determination of the gluon density in the proton was carried out in DIS and in the photoproduction regime. The result 
agrees with the indirect determinations from scaling violations.

Inelastic \jpsi\ production is studied in photoproduction and DIS and 
is well described in photoproduction by the \colsingmodel\ alone in \nlo. 
In DIS the data have been compared to LO \colsing\ and \coloct\ predictions 
(at parton level). In the latter rough agreement in absolute normalisation 
is found, while the \colsing\ model reproduces the shape of the data slightly 
better.
 
\vspace{1cm}
\paragraph*{Acknowledgement}
I wish to thank the organisers for a very pleasant and fruitful meeting and 
my colleagues at ZEUS and H1 for supplying their data and for discussions.

\end{document}